\documentclass[sigconf]{acmart}
\usepackage{tabularx}
\usepackage[inkscapelatex=false]{svg}
\AtBeginDocument{%
  }

\setcopyright{none}
\acmDOI{}
\acmConference[LOCO '26]{2nd International Workshop on Low Carbon Computing}{10--11 September 2026}{Lancaster, UK}
\acmISBN{}

\begin{document}

\title[Could Model Partitioning Make Federated Learning More Sustainable?]{Could Model Partitioning Make Federated Learning\\More Sustainable?}

\author{Tobias Frohlich}
\affiliation{%
  \institution{University of Glasgow}
  \city{Glasgow}
  \country{United Kingdom}
}
\email{t.frohlich.1@research.gla.ac.uk}

\author{Tiffany Vlaar}
\affiliation{%
  \institution{University of Glasgow}
  \city{Glasgow}
  \country{United Kingdom}
}
\email{tiffany.vlaar@glasgow.ac.uk}

\author{Lauritz Thamsen}
\affiliation{%
  \institution{University of Glasgow}
  \city{Glasgow}
  \country{United Kingdom}
}
\email{lauritz.thamsen@glasgow.ac.uk}

\renewcommand{\shortauthors}{Frohlich et al.}

\begin{abstract}
As federated learning (FL) extends from distributed machine learning between low-power devices to cross-silo scenarios involving edge servers and data centres, its carbon footprint has become a growing concern.
Addressing this, methods for sustainable FL align training with low-carbon energy availability or low grid demand and reduce the energy consumption of clients powered by high-carbon sources by decreasing the size of their models.

We propose applying model partitioning, which can shift energy consumption by offloading parts of a model to another participant, in response to carbon- or grid-aware signals.
Our preliminary findings show that for some partition points, model partitioning can reduce a participant's energy consumption by up to 76\% without any significant time or energy consumption overhead compared to non-partitioned training.
\end{abstract}


\maketitle

\section{Introduction}
In federated learning (FL), neural networks are trained collaboratively across a distributed network of devices, where each device trains a local model with its own data before these local models are aggregated into a global model at a central server.
Originally applied to cross-device settings that feature a large number of low-power clients, such as smartphones or IoT devices, today's use cases for FL extend to cross-silo scenarios with fewer but more powerful clients at an edge server or even data centre scale~\cite{bian_cafe_2024}.

This development is concerning from a sustainability perspective. The carbon footprint of machine learning on energy-intensive hardware~\cite{strubell_energy_2019, wu_sustainable_2022, luccioni_estimating_2023} threatens to exacerbate the carbon emissions of the Information and Communication Technology (ICT) sector, which must decrease significantly in order to meet climate targets~\cite{freitag_real_2021}. 
Although processing large amounts of data close to the source rather than transferring them to a data centre for centralised learning can have sustainability and privacy benefits, FL does not automatically result in a low carbon footprint, even when training on low-power hardware~\cite{wu_sustainable_2022}.
Therefore, as larger models are being trained federated on increasingly more powerful hardware, the need for sustainable FL is more urgent than ever.

In this work, we propose considering model partitioning, where neural networks are split and distributed across multiple devices, as a method for making FL more sustainable.
Typically, models are split between the clients and a more powerful server, with the first few blocks on the clients and the later blocks on the server, to reduce the computational load on the clients without affecting the final model size~\cite{gupta_distributed_2018, thapa_splitfed_2022}. 
Furthermore, the model distribution can be adapted dynamically to fulfil optimisation objectives, such as minimising energy consumption~\cite{guo_hierarchical_2024, samikwa_dfl_2024}.

Model partitioning has not yet been employed to optimise the use of low-carbon energy, for example from on-site renewable sources or via carbon intensity or demand response grid signals. 

This paper makes the following contributions:
\begin{itemize}
    \item We propose that adapting clients' energy consumption through model partitioning in response to the availability of low-carbon energy could make FL more sustainable. 
    \item Using our custom model partitioning framework\footnote{Available at \href{https://github.com/GlasgowC3lab/model_partitioning_loco2026}{https://github.com/GlasgowC3lab/model\_partitioning\_loco2026}}, we conduct preliminary experiments to show how partitioning can shift energy consumption between client and server GPUs for different model architectures and partition points.
    \item We present three examples that illustrate the potential of using model partitioning for signal-based shifting.
\end{itemize}

\section{Background}
\subsection{Sustainability Signals}
The operational carbon emissions of a computing process are determined by its energy consumption and the carbon intensity of the energy source, i.e., the amount of carbon emitted by producing one unit of electricity.
Renewable energy sources, such as wind or solar, have significantly lower carbon intensities than fossil fuel-based sources, such as gas or coal.

This has given rise to the concept of \textit{carbon awareness} in computing: aligning the execution of workloads with the availability of low-carbon energy through temporal and geospatial load shifting.
The energy source dictates which signal workloads can be aligned with.
For hardware powered by on-site renewables, the amount of available energy depends on the current production of these renewables, which is often subject to weather conditions. Energy production forecasts can be used as a signal~\cite{wiesner_fedzero_2024}.
When power is instead supplied by regional electricity grids, workloads can be aligned with the grid's average carbon intensity~\cite{bian_cafe_2024}, i.e., the average carbon intensities of all energy sources weighted by their contribution to the current grid mix.

However, average carbon intensity ignores the effect of demand on the electricity grid mix. When grid demand is higher than renewable supply, shifting workloads to these periods might require fossil fuel power plants to increase production to meet the demand, which would reverse any potential carbon savings achieved by the shift. Therefore, marginal carbon intensity instead considers the carbon intensity of the energy that would be generated to fulfil the increased grid demand. These signals are not always accurate, as determining the marginal generator can be difficult in real grids. Furthermore, while marginal carbon intensity can be a sensible signal for online decision making, average carbon intensity is commonly used for carbon accounting.

Many grids further offer signals for grid demand in the form of demand response programs (DRPs), which are schemes to incentivise reducing and increasing electricity use during peak and off-peak periods, respectively. 
These programs aim to stabilise grids and can increase their sustainability, as they can prevent increased fossil fuel-based energy generation during peak demand and renewable energy curtailment due to low demand~\cite{sousa_effect_2023}.
    
FL is well-suited for carbon and grid awareness since federations often span across multiple different electricity grids, client participation can change from epoch to epoch, and deadlines are usually flexible. At the same time, FL systems can be large-scale, involving considerable training data and sizeable models, resulting in significant energy consumption and emissions.

\subsection{Model Partitioning}
In this paper, we refer to model partitioning as splitting neural networks into blocks consisting of one or more layers and then distributing these over multiple devices.
The client typically retains the first blocks of the model and performs the forward pass for its data before sending activations and labels to the server, which holds the rest of the model.
The server finishes the forward pass with the received activations and calculates the gradients, which are then transmitted back to the client to finish backpropagation. After training concludes, the client- and server-side parts are combined into a full-size model. Therefore, model partitioning can reduce the computational load on clients without altering the size of the final model.

Federated training with model partitioning is also referred to as \textit{split learning}.
While early versions of split learning were sequential~\cite{gupta_distributed_2018}, this process was later parallelised in the form of \textit{split federated learning}, where clients train in parallel and models are then aggregated~\cite{thapa_splitfed_2022}.
As layer types differ in computational complexity and activation size, the optimal distribution of blocks between client and server depends on the devices and the communication capabilities between them~\cite{samikwa_dfl_2024}.

\section{Model Partitioning for Sustainable FL}
We envision using model partitioning to reduce FL's carbon emissions by responding to local electricity grid conditions and on-site energy availability, thereby optimising the use of low-carbon energy.

Using model partitioning to offload parts of the model training to other participants can enable shifting energy consumption towards the availability of low-carbon energy or low grid demand, which could contribute to balancing supply and demand in electricity grids.
Offloading work instead of reducing participation where low-carbon energy is unavailable increases fairness, as it allows clients to contribute their data regardless of sustainability constraints. The previously discussed carbon- and grid-aware signals can be used to identify low-carbon energy availability or low grid demand.
    
In this work, we investigate how much energy consumption model partitioning can shift between large-scale hardware to assess the feasibility of this vision.

\subsection{Experiment Setup}
We conduct a preliminary experiment to explore how training a partitioned model between client and server affects the individual and total GPU energy consumption under FL training conditions.
We consider models with two different architectures split at several different points between client and server. 

We train client-side models on an NVIDIA L40S GPU and server-side models on an NVIDIA H100 NVL GPU, measuring their power consumption with nvidia-smi. We evaluate the 50-, 101-, and 152-layer versions of ResNet~\cite{he_deep_2016} and the 11- and 19-layer variants of VGG~\cite{simonyan_very_2015}, two image classification models widely used in model partitioning experiments. ResNet-50, ResNet-101, and ResNet-152 were partitioned into 8 blocks of one or more layers and the significantly more shallow VGG-11 and VGG-19 were partitioned into 7 blocks, as shown in Table \ref{tab:partition-points}. The partition points were chosen manually based on the model architecture, for example, after self-contained blocks or pooling layers that reduce the activation size.

Each model is trained on CIFAR-10~\cite{krizhevsky_learning_2009} for 50 global epochs for all block distributions. Our baseline, denoted as $\star$ in the figures and tables, is partition point $p$ = 8 for ResNet and $p$ = 7 for VGG, where the full model is trained non-partitioned on the client as in vanilla FL. We use the SGD optimiser with batch size 128, learning rate 0.001, and a momentum of 0.9, as these hyperparameters led to well-performing models during tuning. All reported results are the average of 3 individual runs.
\begin{table}
  \caption{Partition point configurations used for the two architectures considered. The partition point $p$ indicates the last block in the client-side model. $\star$ indicates the full model.}
  \label{tab:partition-points}
  \begin{tabularx}{\linewidth}{cXX}
    \toprule
    $p$&ResNet&VGG\\
    \midrule
    1 & Conv & Conv Block (multiple Conv + MaxPool)\\
    2 & MaxPool & Conv Block\\
    3 - 5 & Bottleneck Block (multiple Conv) & Conv Block\\
    6 & Bottleneck Block & Linear \\
    7 & AvgPool & 2 Linear $\star$\\
    8 & Linear $\star$ & - \\
  \bottomrule
\end{tabularx}
\end{table}

\subsection{Preliminary Results}
\begin{table}
  \caption{Best test accuracy (\%) for all model configurations. $\star$ denotes the non-partitioned FL baseline.}
  \label{tab:accuracy}
  \begin{tabular}{cccccc}
    \toprule
    $p$&ResNet-50&ResNet-101&ResNet-152&VGG-11&VGG-19\\
    \midrule
    1 & 90.2 & 90.0 & 88.8 & 82.2 & 85.2\\
    2 & 90.2 & 90.0 & 88.8 & 82.6 & 85.8\\
    3 & 89.6 & 89.8 & 89.4 & 83.4 & 85.2\\
    4 & 90.8 & 88.8 & 89.6 & 84.6 & 87.0\\
    5 & 88.2 & 89.8 & 90.2 & 85.4 & 89.6\\
    6 & 86.8 & 87.8 & 91.2 & 87.0 & 91.4\\
    7 & 86.8 & 87.8 & 91.2 & 86.2 $\star$ & 90.4 $\star$\\
    8 & 86.0 $\star$ & 87.6 $\star$ & 88.2 $\star$ & - & -\\
  \bottomrule
\end{tabular}
\end{table}

\begin{figure}[h]
  \centering
  \includegraphics[width=\linewidth]{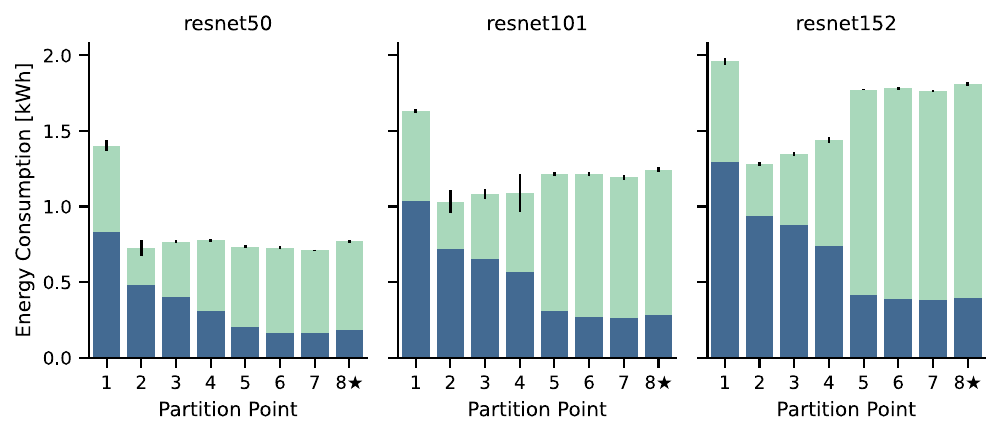}
  \includegraphics[width=\linewidth]{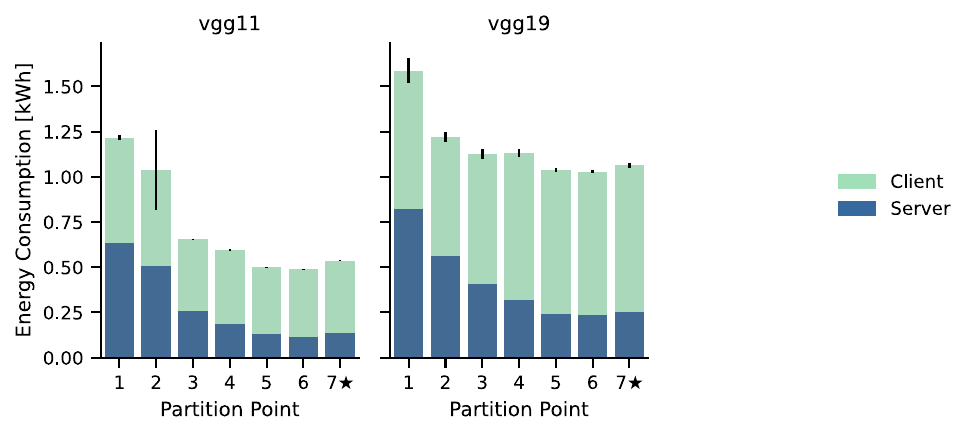}
  \caption{GPU energy consumption of client and server for all model configurations. $\star$ denotes the non-partitioned FL baseline.}
  \label{fig:energy}
\end{figure}

\begin{figure}[h]
  \centering
  \includegraphics[width=\linewidth]{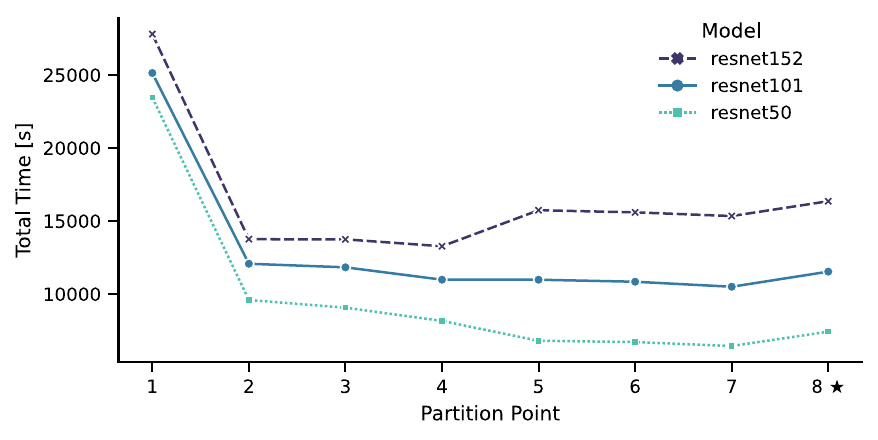}
  \caption{Total time for all ResNet configurations. $\star$ denotes the non-partitioned FL baseline.}
  \label{fig:time}
\end{figure}

\paragraph{ResNet}
Figure \ref{fig:energy} shows the GPU energy consumption at client and server with different partition points. For later partition points ($p$ = 5, 6, and 7), the combined energy consumption of client and server GPUs is similar to full training on the client. For earlier partition points ($p$ = 2, 3, and 4), the client consumed less energy but the server more, so some energy consumption was essentially shifted from the client to the server. Client energy savings also scale with the model size.
For example, partitioning the model at $p$ = 2 results in a 58\% reduction in client energy consumption for ResNet-50, 68\% for ResNet-101, and 76\% for ResNet-152 compared to full on-client training.
The overall energy use is consistent across all ResNet configurations, suggesting that energy consumption was shifted without increasing it. 
For ResNet-152, partitioning at early blocks even reduced the overall energy consumption by up to 29\% ($p$ = 2) compared to full model execution, since a large portion of the model was then trained on the more energy-efficient server GPU.

Total time was similar across ResNet configurations, although training took longer for earlier partition points for some models (Figure \ref{fig:time}).
ResNet-50 saw an increase in total time of up to 29\% ($p$ = 2), although this decreases to 15\% ($p$ = 2) for ResNet-101 and no such effect is evident for ResNet-152, where full model training took longer than all partitioned configurations and earlier partition points resulted in time reductions of up to 19\% ($p$ = 4).

Overall, the results indicate a shifting overhead for $p$ = 1. Although client energy consumption was reduced compared to the baseline for all models, the total time and energy consumption were significantly larger than for all other configurations.

As shown in Table \ref{tab:accuracy}, ResNet's generalisation performance was not negatively affected by partitioning, as the test accuracy was lowest for $p$ = 8 and increased up to 4.8\% for ResNet-50 ($p$ = 4), 2.4\% for ResNet-101 ($p$ = 1 and 2), and 3.0\% for ResNet-152 ($p$ = 6 and 7) when partitioned. 

\paragraph{VGG}
For VGG-11, partitioning achieved only minor reductions in client energy consumption of up to 7\% ($p$ = 5), as shown in Figure \ref{fig:energy}. VGG-19 saw client energy consumption decrease by up to 19\% ($p$ = 2), although the total energy consumption slightly increased for the configurations with the most client energy savings. 

Both VGG models had lower total training times for later partition points ($p$ = 5 and 6) but early partition points showed an increase of up to 204\% for VGG-11 and 56\% for VGG-19 ($p$ = 2).

As for ResNet, $p$ = 1 also had the highest energy consumption and total time for both VGG models.

VGG-11 and VGG-19's test accuracy decreased up to 4.0\% ($p$ = 1) and 5.2\% ($p$ = 1 and 3), respectively, for early partition points but improved slightly for $p$ = 6 compared to the baseline (Table \ref{tab:accuracy}).

\paragraph{Discussion}
The high total time and energy consumption at $p$ = 1 might be caused by communication overheads, as we estimated a considerable data transmission overhead for most partitioned configurations, which was largest for early partition points (150-400$\times$ that of the FL baseline for $p$ = 1 compared with 1-2.5$\times$ for $p$ = 6 (VGG) and $p$ = 7 (ResNet)). However, how this data overhead translates to energy consumption depends on multiple factors, including the network path and conditions as well as the exact energy attribution method. Given the highly static power draw of core networks~\cite{myttonNetworkEnergyUse2024}, the energy overhead could be considerably less than the data overhead, and further work is required to quantify and compare communication and training energy costs.

Both ResNet and VGG saw variations in test accuracy between different partition points. Such fluctuations can likely be ascribed to PyTorch implementation details, as previously observed by~\cite{dachilleImpactCutLayer2025}, and should be studied further.
 
Overall, the effectiveness and overhead of shifting depend substantially on the partition point. As the optimal partition point varies from one model to the next, careful selection of the partition point is crucial. VGG further benefitted less from partitioning than ResNet, presumably due to architectural differences. The more shallow VGG models, which can be trained efficiently in full on the client GPU, gained little speed-up from offloading to the more performant server GPU but might have been more strongly affected by the communication overhead, which increased the total time and energy consumption for earlier partition points.

\subsection{Signal-Based Shifting Potential}
We briefly describe three scenarios that exemplify the potential of using model partitioning in response to sustainability signals under the assumption of sufficient client- and server-side node availability.

\paragraph{On-Site Renewable Energy}
We consider the partitioned training of ResNet-101 between a cloud server and a client which is powered by on-site solar energy with 0.45 kWh available for the training duration. Our experimental results suggest that partitioning at $p$ = 3 and offloading later blocks to the server could reduce the client’s energy consumption from 0.96 kWh to 0.43 kWh, ensuring that the client almost completely uses its on-site renewable energy. Changing the partition point during training might further enable a more fine-grained adaptation of the client’s energy consumption to optimally use all of the renewable energy available, and we leave the exploration of this idea for future work.

\paragraph{Grid Carbon Intensity}
We assume static average carbon intensities $CI$ for training ResNet-152 between a London cloud server ($CI$ = 172 gCO\textsubscript{2}/kWh) and different clients across the UK, comparing the partition point that led to the largest emission reductions with the baseline of full model training on the clients\footnote{Carbon intensities retrieved from \href{https://carbonintensity.org.uk/}{https://carbonintensity.org.uk/}}. 
For clients with access to very low-carbon energy, where the overall emissions are dominated by the server, partitioning would only result in minor savings. For example, for a client in South Scotland ($CI$ = 2 gCO\textsubscript{2}/kWh), offloading could reduce total emissions slightly from 71.0 gCO\textsubscript{2} to 68.5 gCO\textsubscript{2} ($p$ = 7). 

Splitting leads to larger benefits when the client carbon intensity is higher, as it could decrease the carbon emissions of a client in Yorkshire ($CI$ = 134 gCO\textsubscript{2}/kWh) from 257.6 gCO\textsubscript{2} to 207.3 gCO\textsubscript{2} ($p$ = 2). For a high-carbon client in South Wales ($CI$ = 341 gCO\textsubscript{2}/kWh), partitioning could cut the overall training carbon emissions in half, saving about 271.8 gCO\textsubscript{2} ($p$ = 2).

We plan to investigate the question of optimal partitioning under variable carbon intensities in subsequent work.

\paragraph{Demand Response Programs}
For this example, we consider a client participating in a DRP with the ability to offload part of its ResNet-152 training to a cloud server. Assuming the DRP signal demands an energy consumption reduction of 60\% for the training period, the client could comply by partitioning at $p$ = 3 or $p$ = 2 for a 67\% or 76\% reduction, respectively.

Coordinating the participation of many widely distributed clients in possibly different DRPs is another avenue for future research.
  
\section{Related Work}
Previous works have proposed different methods to increase the sustainability of FL.
Most methods focus on client selection and scheduling based on different sustainability signals, such as the availability of on-site renewables~\cite{wiesner_fedzero_2024}, grid carbon intensity~\cite{bian_cafe_2024}, or DRPs~\cite{wang_toward_2023}.
Our work aligns more closely with methods that use sustainability signals to dynamically reduce clients' workload through quantisation~\cite{wang_toward_2023}, pruning~\cite{li_fedcarbon_2024}, or ordered dropout~\cite{abbasi_fedgreen_2024}, yet model partitioning has not been used as a method for this before.

Existing works on model partitioning have explored dynamically adapting the layer distribution to fulfil certain optimisation criteria, such as minimising energy consumption~\cite{samikwa_dfl_2024, guo_hierarchical_2024}.
Previous investigations of model partitioning from a sustainability perspective do not consider carbon or grid signals~\cite{guo_hierarchical_2024} or are limited to inference~\cite{ke_carboncp_2024}, whereas we explore combining model partitioning with sustainability signals for FL training.

\section{Conclusion and Future Work}
We believe that using model partitioning to shift energy consumption within federations in response to carbon- or grid-aware signals, such as carbon intensity, DRPs, or renewables availability, could be a tool to make FL training more sustainable in the light of rising ICT sector carbon emissions. 
Our preliminary findings suggest that this shift does not have to affect the learning performance or total energy consumption for some partition points, although additional experiments are needed with multiple clients as well as other models, particularly those associated with substantial energy consumption, such as large language models, and heterogeneous hardware with less powerful GPUs, such as edge or consumer devices.

In the future, we plan to evaluate partitioning in response to different carbon- and grid-aware signals and to automatically determine optimal partition points in dynamically changing systems, which we expect to involve efficiently profiling and modelling the impact of different block distributions on heterogeneous hardware and under different network conditions.
Furthermore, we intend to explore decentralised offloading scenarios to optimally exploit variability within geo-distributed FL systems.

\begin{acks}
This work was supported by the EPSRC Centre for Doctoral Training in Diversity-Led, Mission-Driven Research (DiveIn CDT), Grant Number EP/Y034902/1.
\end{acks}

\section*{Rights Retention}
For the purpose of open access, we have applied a Creative Commons Attribution (CC BY) licence to this manuscript.

\section*{Data/Code Availability}
An open-source implementation of our model partitioning framework is available at \href{https://github.com/GlasgowC3lab/model_partitioning_loco2026}{https://github.com/GlasgowC3lab/model\_ partitioning\_loco2026}.

\bibliographystyle{ACM-Reference-Format}
\bibliography{references}

\end{document}